# A Solution to the Strong CP Problem

G. Schierholz

Deutsches Elektronen-Synchrotron DESY, D–22603 Hamburg, Germany
and
Gruppe Theorie der Elementarteilchen, Höchstleistungsrechenzentrum HLRZ,
c/o Forschungszentrum Jülich, D–52425 Jülich, Germany

**Abstract**

One may argue that QCD solves the strong CP problem by itself, without having to introduce new symmetries and particles. To test this idea, a lattice simulation is performed. The problem is investigated in the $CP^3$ model first. It is found that the model has a first order phase transition in $\theta$ from a confining phase at small $\theta$ to a deconfining phase at large $\theta$, and that the critical value of $\theta$ decreases towards zero as $\beta$ is taken to infinity. This suggests that $\theta$ is tuned to zero in the continuum limit. Preliminary studies of the SU(2) Yang-Mills theory in four dimensions show a phase transition in $\theta$ as well, so that it is quite likely that the strong CP problem in QCD is solved along the same line.

In QCD, as well as in other theories that possess instantons, the proper vacuum states are superpositions of vacua of different winding numbers $n$:

$$|\theta\rangle = \sum_n \exp(\mathrm{i}\theta n)|n\rangle, \theta \in [0, 2\pi). \tag{1}$$

These so-called $\theta$ vacua are realized by adding a CP violating term to the action,

$$S_\theta = S - \mathrm{i}\theta Q, \tag{2}$$

where $S$ is the standard action and $Q$ is the topological charge. A priori $\theta$ is a free parameter. Since no CP violation has been observed in the strong interactions, $\theta$ must however be very close to zero. † This constitutes the strong CP problem. The present upper bound is $\theta \leq 10^{-9}$ [1].

A popular picture of the QCD vacuum is that of a dual superconductor [2] in which color magnetic monopoles condense and color electric charges, i.e. quarks and gluons, are confined by a dual Meissner effect. This picture has been successfully tested in lattice simulations [3]. In the $\theta$ vacuum these monopoles acquire a color electric charge of the magnitude $\theta/2\pi$ [4]. For $\theta \neq 0$ one would expect that the long-range color electric forces are screened by monopoles and that confinement is lost. So QCD would only be a viable (continuum) theory for $\theta = 0$.

What makes a lattice simulation at non-vanishing values of $\theta$ very difficult is the fact that the action is complex. As a result, standard lattice techniques are not immediately applicable. This has led us to investigate the problem in a simpler model first.

A model, which in many respects is similar to QCD, is the $CP^{N-1}$ model in two space-time dimensions. The $CP^{N-1}$ model deals with N-component, complex scalar fields $z_a(x)$ of unit length: $\overline{z}_a(x)z_a(x) = 1$, $a = 1, \cdots, \mathrm{N}$. Out of these fields one constructs composite vector fields

$$A_\mu(x) = \frac{\mathrm{i}}{2}\overline{z}_a(x) \overleftrightarrow{\partial}_\mu z_a(x). \tag{3}$$

† In the presence of quarks the $\theta$ angle can be rotated into a phase of the quark mass matrix $M$ and vice versa, due to the $U_A(1)$ anomaly. The effective angle is in this case $\bar{\theta} = \theta + \arg\det M$.

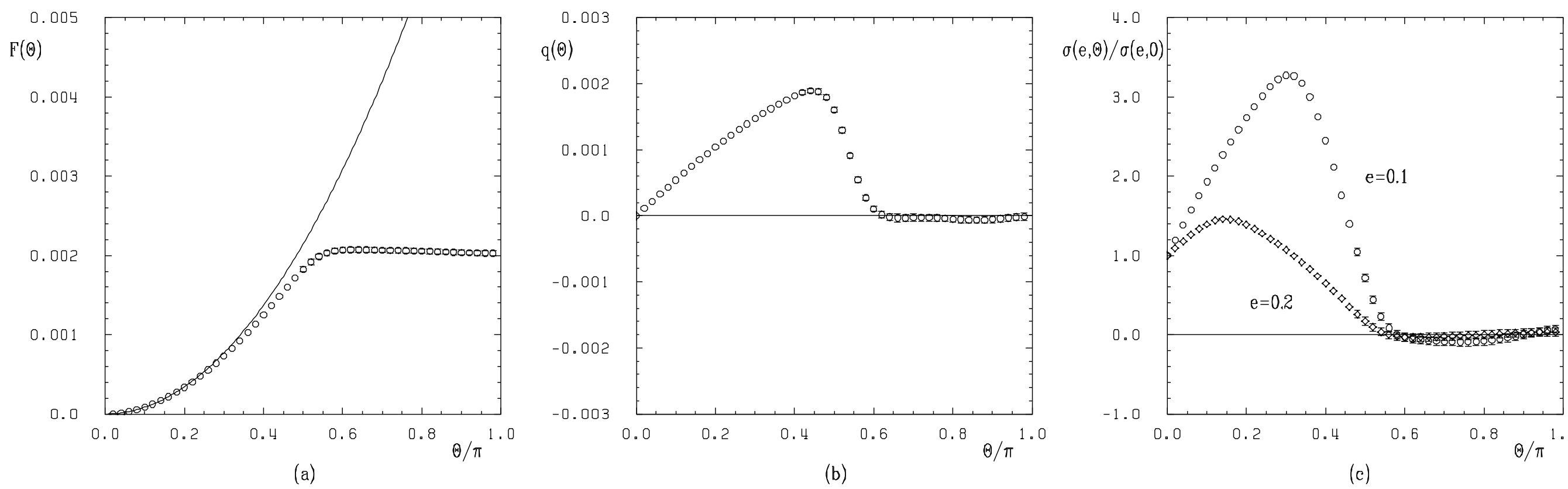

**Figure 1.** The free energy $F(\theta)$ (a), the charge density $q(\theta)$ (b) and the string tension $\sigma(e,\theta)$ (c) as a function of $\theta$ on the $V = 64^2$ lattice at $\beta = 2.7$. The solid curve in (a) is the prediction of the large-N expansion to leading order. Only the first half of the $\theta$ interval is displayed. In the second half of the interval $F(\theta) = F(2\pi - \theta)$.

Then the action [5] can be written [6]

$$S = \beta \int d^2x \overline{D_\mu z}_a(x) D_\mu z_a(x), \qquad (4)$$

where $D_\mu = \partial_\mu + \mathrm{i}A_\mu$. Thus the model describes a set of charged scalar fields interacting minimally with a composite gauge field. The topological charge is given by

$$Q = \frac{1}{2\pi} \int d^2x \epsilon_{\mu\nu} \partial_\mu A_\nu(x) \equiv \frac{1}{2\pi} \int d^2x F_{01}. \qquad (5)$$

The $\theta$ dependence of the theory is governed by the partition function

$$Z(\theta) = \sum_Q \exp(\mathrm{i}\theta Q)\, p(Q) \equiv \exp(-VF(\theta)), \qquad (6)$$

where $p(Q)$ is the probability of finding a field configuration with charge $Q$, and $F(\theta)$ is the free energy per space-time volume $V$. In terms of $F(\theta)$ the average topological charge density is given by

$$\frac{1}{V}\langle\theta|Q|\theta\rangle \equiv -\mathrm{i}q(\theta) = -\mathrm{i}\frac{dF(\theta)}{d\theta}, \qquad (7)$$

and the string tension of two external particles of charge $e$ and $-e$ (in units of the intrinsic charge) turns out to be

$$\sigma(e,\theta) = F(\theta + 2\pi e) - F(\theta). \qquad (8)$$

We have chosen to investigate the $\mathrm{CP}^3$ model. For details of the calculation see ref. [7, 8].

If our idea is correct, we should find a first order phase transition in $\theta$ from a confining phase to a Higgs or Coulomb phase. On a finite lattice and at a finite value of $\beta$ the phase transition is expected to occur at a value $\theta = \theta_c(\beta, V) > 0$, where $V$ now is the lattice volume. Only on an infinite lattice and at $\beta = \infty$, i.e. in the continuum limit, would we expect that $\theta_c = 0$.

A first order phase transition manifests itself in a kink in the free energy, as well as in a discontinuity in the first derivative of the free energy, i.e. the topological charge density. In Fig. 1 I show $F(\theta)$ on the $V = 64^2$ lattice at $\beta = 2.7$. For comparison I also show the prediction of the large-N expansion to leading order [6], $F(\theta) = \text{const.} \cdot \theta^2$. We see a distinctly marked kink at $\theta = \theta_c \approx 0.5\,\pi$: while $F(\theta)$ increases roughly proportional to $\theta^2$ up to $\theta = \theta_c$, $F(\theta)$ turns out to be constant (within the error bars) for $\theta \geq \theta_c$. In Fig. 1 I also show the topological charge density on the same lattice. According to (6) this can be interpreted as a background electric field. We see that $q(\theta)$ increases almost linearly with $\theta$ up to $\theta = \theta_c$, where it jumps to zero and then stays zero over the rest of the interval (again within the error bars). Thus the phase transition is marked by a collapse of the background electric field, presumably due to pair production.

To show that the phase transition is a deconfining phase transition, I have plotted the string tension in Fig. 1, again on the $64^2$ lattice at $\beta = 2.7$, for two different charges. To keep the discussion simple, I have restricted myself to the case of small fractional charges. This matters because the vacuum will change its properties if it is exposed to a strong external electric field. We see that the string tension is zero (within the error bars) for $\theta \geq \theta_c$. This result is a consequence of the property that $F(\theta)$ is constant for $\theta \geq \theta_c$. The string tension reaches its peak at $\theta = \theta_c - 2\pi e$.

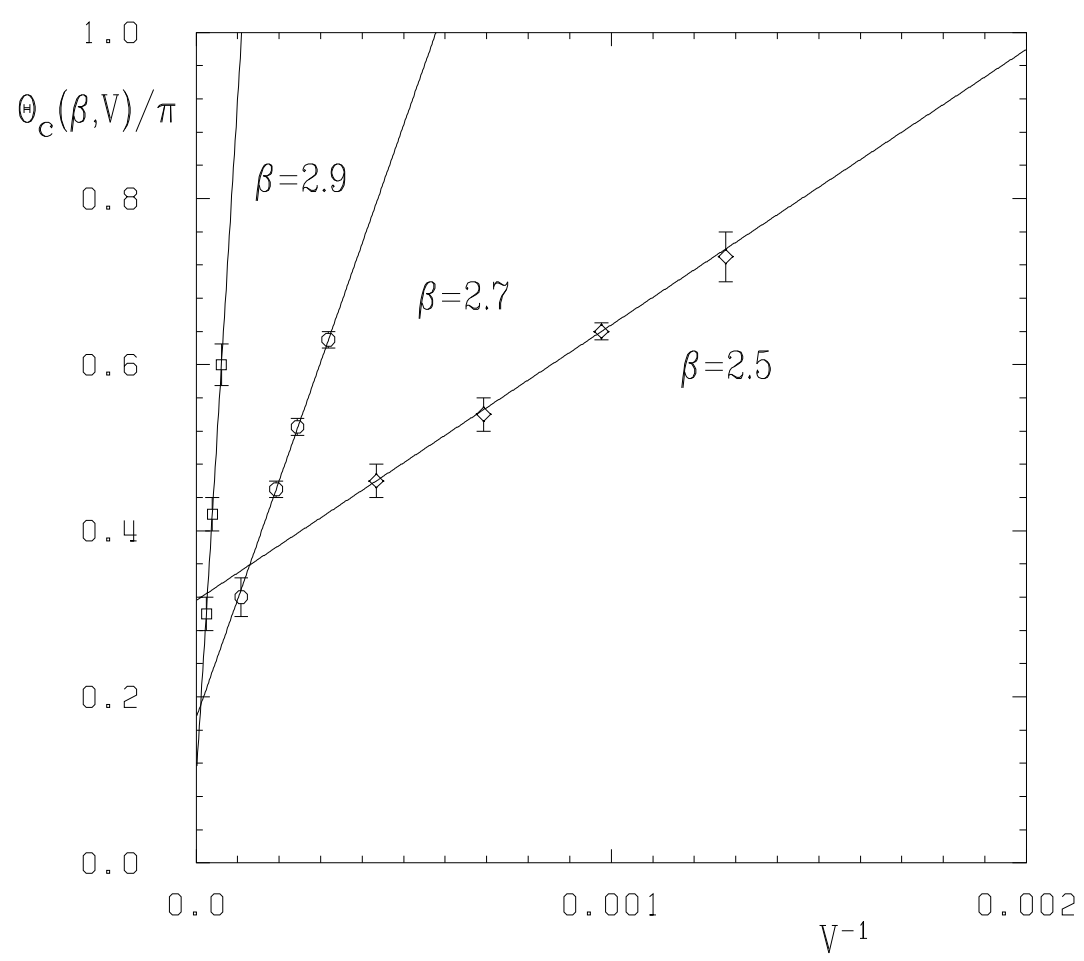

**Figure 2.** The critical value $\theta_c(\beta, V)$ as a function of $V^{-1}$ for three values of $\beta$. The lines are a linear fit to the data points.

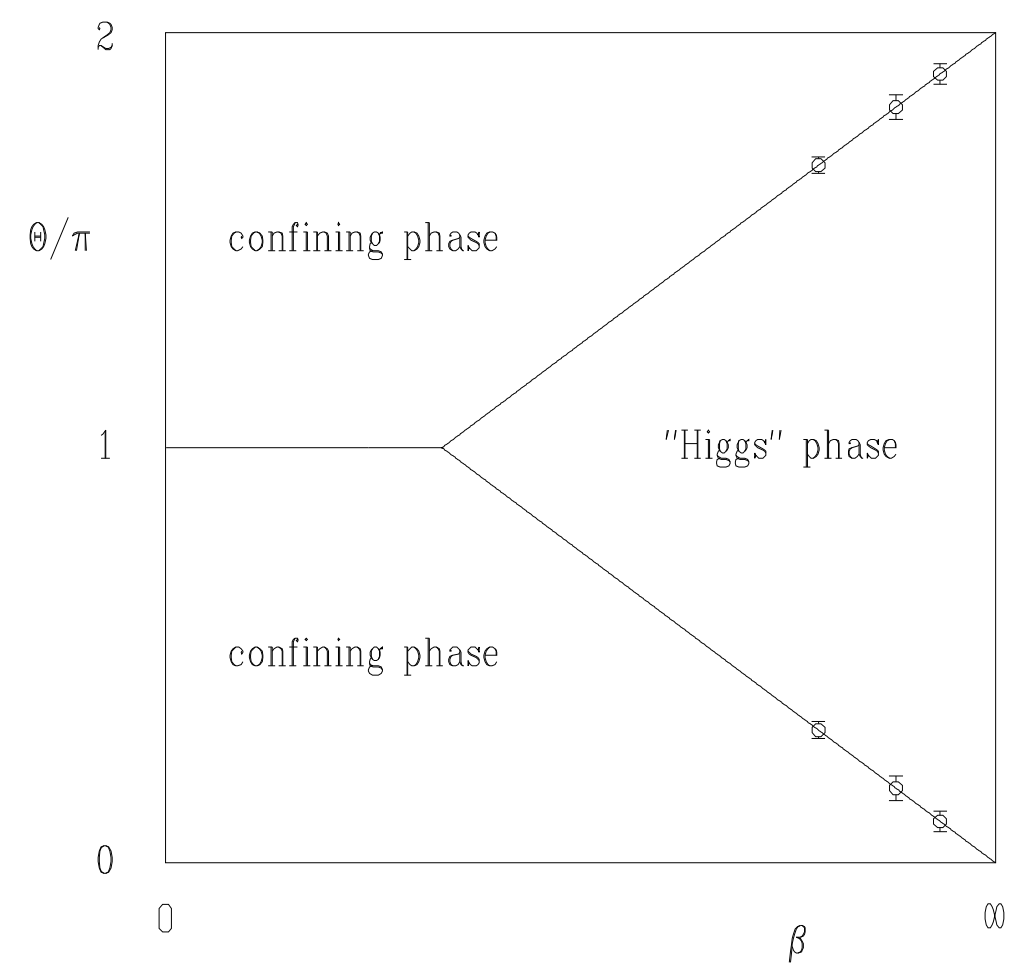

**Figure 3.** The phase diagram. The horizontal line at $\theta = \pi$ is the strong coupling prediction, and the symbols are the lattice data extrapolated to infinite volume.

In Fig. 2 I show all our results for $\theta_c(\beta, V)$ for a variety of lattice volumes ranging from $V = 28^2$ to $V = 200^2$ and for three values of $\beta$: $\beta = 2.5$, $\beta = 2.7$ and $\beta = 2.9$. The lattice data display strong finite size effects. For a first order phase transition we expect

$$\theta_c(\beta, V) - \theta_c(\beta, \infty) \propto V^{-1} \qquad (9)$$

for fixed $\beta$. When $\theta_c(\beta, V)$ is plotted against $V^{-1}$, we find that for all three values of $\beta$ our data fall on a straight line, in accordance with the predictions of a first order phase transition. This allows us to extrapolate the lattice results to the infinite volume.

We find that the extrapolated values of $\theta_c$ fall rapidly to zero as $\beta$ is taken to infinity. Our results are consistent with a decay like $\theta_c \propto 1/\xi$, where $\xi$ is the correlation length. (For our values of $\beta$ the correlation length is 4.5, 8.8 and 18.5, respectively.) In the strong coupling limit the model is known to have a first order phase transition at $\theta = \theta_c = \pi$ [9]. This suggests a phase diagram of the form shown in Fig. 3.

Thus $\theta = 0$ is the only value at which one can take the continuum limit, at least within the confining phase. This resolves the strong CP problem.

As a next step towards solving QCD, we are currently investigating the problem in the SU(2) Yang-Mills theory in four dimensions [10]. Preliminary results indicate a first order phase transition in $\theta$, just as in the case of the $CP^3$ model. Because of lack of space I cannot show a figure here. To establish the phase diagram firmly, we will have to repeat the calculation on lattices of various sizes and for several values of $\beta$ still, as we did for the $CP^3$ model.

[1] V. Baluni, Phys. Rev. **D19** (1979) 2227;
R. Crewther, P. Di Vecchia, G. Veneziano and E. Witten, Phys. Lett. **B88** (1979) 123; Erratum, *ibid.* **B91** (1980) 487.
[2] G. 't Hooft, in *High Energy Physics*, Proceedings of the EPS International Conference on High Energy Physics, Palermo, 1975, ed. A. Zichichi (Editrice Compositori, Bologna, 1976);
G. 't Hooft, Phys. Scripta **25** (1982) 133;
S. Mandelstam, Phys. Rep. **C23** (1976) 245.
[3] A. S. Kronfeld, G. Schierholz and U.-J. Wiese, Nucl. Phys. **B293** (1987) 461;
A. S. Kronfeld, M. L. Laursen, G. Schierholz and U.-J. Wiese, Phys. Lett. **B198** (1987) 516;
F. Brandstaeter, G. Schierholz and U.-J. Wiese, Phys. Lett. **B272** (1991) 319;
V. Singh, D. A. Browne and R. W. Haymaker, Phys. Lett. **B306** (1993) 115;
for a recent review see:
T. Suzuki, Nucl. Phys. **B** (Proc. Suppl.) **30** (1993) 176, and references therein.
[4] E. Witten, Phys. Lett. **B86** (1979) 283.
[5] H. Eichenherr, Nucl. Phys **B146** (1978) 215;
E. Cremmer and Scherk, Phys. Lett. **B74** (1978) 341.
[6] A. D'Adda, P. Di Vecchia and M. Lüscher, Nucl. Phys. **B146** (1978) 63;
E. Witten, Nucl. Phys. **B149** (1979) 285.
[7] Š. Olejník and G. Schierholz, Nucl. Phys. **B** (Proc. Suppl.) **34** (1994) 709.
[8] G. Schierholz, Nucl. Phys. **B** (Proc. Suppl.) **37A** (1994) 203.
[9] N. Seiberg, Phys. Rev. Lett. **53** (1984) 637.
[10] F. Brandstaeter and G. Schierholz, in progress.